\documentclass[prl,twocolumn,floatfix,showpacs]{revtex4}
\usepackage{psfig}
\begin{document}
\title{Critical Current Peaks at $3B_{\Phi}$ in Superconductors
with Columnar Defects: Recrystalizing the Interstitial Glass}
\author{M. E. Gallamore, G. E. D. McCormack, and 
T. P. Devereaux}
\affiliation{Department of Physics,
University of Waterloo, Waterloo, Ontario Canada N2L 3G1}
\date{\today}
\begin{abstract}
The role of commensurability and the interplay of correlated
disorder and interactions on vortex dynamics in the presence of
columnar pins is studied via molecular dynamics simulations.
Simulations of dynamics reveal substantial caging effects and a
non-monotonic dependence of the critical current with enhancements
near integer values of the matching field $B_{\phi}$ and $3B_{\phi}$ in
agreement with experiments on the cuprates. 
We find qualitative differences in the phase diagram for small and
large values of the matching field.
\end{abstract}
 \maketitle

The rich phases of vortex matter in high temperature
superconductors result from a complex interplay between disorder,
interactions and fluctuations\cite{review}. Understanding the
physics of these competing effects has been a challenge which
underlies the nature of various static and dynamic phases and
transitions between them. Of particular interest is the role of
columnar defects created by
heavy ion radiation in forming more effectively
pinned phases of interacting vortices. 
Extended or correlated disorder is much more effective at pinning
vortices than uncorrelated (point-like) disorder and produces
upward shifts of several tesla in the irreversibility lines and
concomitantly larger critical currents.

The physics of 3D vortex pinning via correlated disorder
can be approximately mapped 
onto the problem of interacting quantum bosons in the
presence of uncorrelated disorder in 2D\cite{Nelson1}. 
This has formed 
the basis for many computational studies via quantum Monte Carlo
simulations of local repulsive bosons\cite{Batrouni} and exact
diagonalization studies\cite{Trivedi}.
Several pinned phases of vortex
matter emerge as a function of $B/B_{\Phi}$, where $B_{\Phi}$ is the
equivalent matching field where the number of vortices equals the
number of defects. For the case where the pins outnumber
the vortices 
$(B< B_{\Phi})$, a Bose-glass phase is formed where
vortices are localized onto columnar defects and
possess an infinite tilt modulus for tilts away from 
columnar alignment. For equal numbers of vortices and defects
$(B=B_{\Phi})$, an analog to a Mott insulating phase exists which
possesses an infinite compression modulus\cite{Nelson1}. The existence
of the Bose-glass phase is well documented\cite{review,Batrouni}, and
evidence for the Mott phase has been found in recent magnetization
relaxation measurements at low temperatures\cite{Beauchamp,Nowak}. 
For the case when
the vortices outnumber the pins, $(B>B_{\Phi})$, the situation is
much less clear. Recent work has suggested that
vortices not accommodated to the columnar
defects are caged by pinned vortices to form a weakly pinned 
``interstitial Bose'' glass. It has been conjectured that 
the melting temperature $T_{m}$ 
decreases rapidly as the density of
interstitial vortices increases for large fields, and either extends smoothly 
into the Bose glass phase\cite{Nelson1} or shows a change in slope 
(kink) at $B=B_{\Phi}$\cite{rad,LV}.
Some experiments\cite{worth,Krusin,Nakielski,Samoilov}
do not show substantial changes of the irreversibility line at $B_{\Phi}$, 
while others show only a mild kink at $B_{\Phi}$
\cite{Beauchamp,Maz,Nowak,Smith99,Smith01}.
The change is even less significant for larger defect concentrations
(larger $B_{\Phi}$). 

Moreover, interesting effects are seen as $B$ further increases past
$B_{\Phi}$. In analogy with the presence of Mott lobes in the dual
superconducting-insulator transition in 3D, one would expect
different phases at commensurate values of vortex filling. Nowak
{\it et al} found that the critical current in 
Tl$_{2}$Ba$_{2}$CuO$_{6+\delta}$ decreases for $B>
B_{\phi}$ but then increases and goes through a maximum near
$3B_{\phi}$\cite{Nowak}, which is currently unaddressed by theory.

One of the weakness of previous numerical studies\cite{Batrouni,Trivedi} 
is the use of a short-range
screened interaction which misses the softening of the shear modulus
in the absence of pins in the weak field limit as noted by 
Fetter, Hohenberg and Pincus\cite{Fetter,Nelson1,NS}. 
The purpose of this paper is to address the importance of this omission by
investigating the properties of the
pinned and moving phases via 3D molecular dynamics simulations of vortices
interacting via a long-range potential in the presence of columnar
defects. Our key result is
that the softening of the shear modulus in the low field 
limit has important implications for the phase diagram.
We confirm the existence of a weakly pinned interstitial glass which
becomes more weakly pinned as the field increases in agreement with previous
studies of short-range interactions\cite{Trivedi}. However
we find a qualitatively new phase diagram for low fields 
where the interstitial glass melts
near $B=B_{\phi}$ and recrystalizes at higher fields when the matching field is
below a critical value. 
In this regime we find strong numerical evidence for enhanced
values of the critical current near three times the matching
field. This questions the quantitative appropriateness
of the mapping of vortex physics onto the physics of the
superconductor-insulator transition without a properly detailed
consideration of long-range interactions.

We model the motion of vortices as coupled pancakes on
neighboring, continuous planes separated by a distance $z$, with
the applied magnetic field aligned perpendicularly to the planes.
The off-lattice simulation models the motion of vortices referenced
by a 2D coordinate ${\bf r}$ under the
influence of pinning and repulsive vortex interactions, and driving, 
line bending, and thermal forces:
\begin{equation}
m_{l}{\bf \ddot{ r_{i}}} = -\eta_{l}{\bf \dot{r_{i}}}+ {\bf
f_{D}+f_{T}} - {\partial H(\{{\bf r_{1},r_{2}...}\})\over{\partial
{\bf r_{i}}}}. \label{eq1}
\end{equation}
Here $f_{D}=\Phi_{0}J/c$ is the Lorentz force per unit length due
to an applied current density $J$ perpendicular to the magnetic
field. 
$\eta_{l}=\frac{\Phi_{0}^{2}}{2\pi\xi^{2}\rho_{n}c^{2}}$ is the
Bardeen-Stephen viscous drag coefficient, with $\Phi_{0}=(hc/2e)$
the flux quantum and $\rho_{n}$ the normal state resistivity. The
Langevin thermal force per unit length $f_{T}$ is normalized to
set the rms vortex velocity via the equipartition theorem. The
Hamiltonian $H$ is the sum of the line tension for bending, and
vortex-vortex and vortex-disorder potentials per unit length constructed 
via London theory. The line tension is given by
$\epsilon_{l}=\epsilon\ln(\kappa)\epsilon_{0}$ with
$\epsilon=(m_{c}/m_{ab})^{2}, \kappa=\lambda/\xi,
\epsilon_{0}=(\Phi_{0}/4\pi\lambda)^{2}$, and $\lambda$ magnetic
penetration depth. The vortex interaction is given by a sum of
pairwise interactions $V_{v-v}({\bf r})=\epsilon_{0}K({\bf
r}/\lambda)$ with $K$ a Hankel function\cite{review}. We model the correlated
extended defects as smooth parabolic traps of width $R_{p}$ and
uniform depth $V_{p}=\epsilon_{l}/4$. Defects are randomly
placed and aligned along the c-axis.

For DC transport,
in most cases the vortex mass per unit length $m_{l}$ 
is overall quite small in comparison with the
other parameters in Eq. (\ref{eq1})\cite{mass} 
and has thus usually been neglected
in previous numerical studies in 3D of vortex dynamics\cite{Reichardt96}.
However, in the case of superconductors near a Mott instability,
such as the underdoped cuprates, the vortex mass can be immensely
enhanced as the system approaches a superconductor-insulator
transition\cite{Doniach}. 
We thus keep this term and simply relate the
vortex mass to the mass of the highly renormalized
electrons confined to the vortex
core region, $m_{l}=m_{eff}n\pi\xi^{2}$, with $n$ the electron density and
$m_{eff}$ the in-plane effective mass of the electron renormalized by strong
interactions.
While we have found that the vortex mass has an impact on the magnitude of
the critical current, the relative $J_{c}$ for different vortex
densities do not drastically depend on the choices made.

Periodic boundary conditions are imposed in the planes to maintain
constant global flux density, and open boundary conditions are
employed along the c-axis.  Temperature is chosen which is high
enough to allow individual vortices to be quickly accommodated to
defects in the absence of a driving current but well below the
glass temperature\cite{Trivedi}.

We measure all energies in units of the bare line bending energy
$\epsilon_{0}$ and measure all lengths in units of $d=4\xi$. A
natural time unit $t_{0}$ is chosen to be
$\pi\kappa\eta_{l}d^{2}/\epsilon_{l}$ and the time step is further
discretized in units of $0.01t_{0}$ for the simulations. The
current is measured in units of the BCS depairing current 
$J_{0}=\Phi_{0}c\kappa/12\sqrt{3}\pi^{2}\lambda^{3}$, and
resistivities in terms of the Bardeen-Stephen flux-flow
resistivity $\rho_{BS}=B\Phi_{0}/(c^{2}\eta_{l})$. 
The other parameters used in the
simulations are dictated by values appropriate for YBa$_{2}$Cu$_{3}$O$_{7}$:
$m_{eff}=5m_{e}, R_{p}=2\xi, \xi=17\AA, z=12\AA, \kappa=100,$ and
$\epsilon=1/25$, giving $H_{c2}=120$T.

Our simulations were performed using up to 40,000 vortex segments in
a $128d\times 128d$ square periodic cell containing 80
planes, where finite size effects were investigated and found to be
minimal. Measurements are taken over an interval of
$10^{6}t_{0}$ after the system has reached a steady state after a
typical time $3\times 10^{5}t_{0}$. We typically average the results over
several hundred realizations of disorder particularly at smaller
driving forces. We measure the average vortex velocity in the
direction of the Lorentz force corresponding to the voltage drop
across the sample, and determine the resistivity $\rho=\mathcal{E}/J$.

\begin{figure}
\psfig{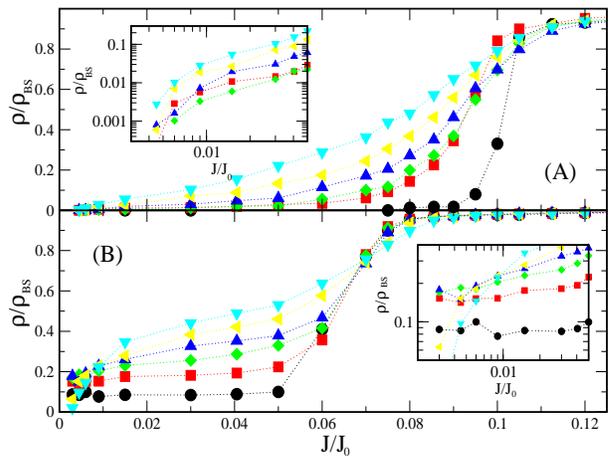} 
\caption[] 
{Resistivity as a function of driving current for different values of the 
applied magnetic field referenced to the matching field $B_{\Phi}$ for
$B_{\Phi}/H_{c2}\times 10^{3}=7.32$ (A) and 1.525 (B), respectively.
The values of $B/B_{\Phi}$ are indicated as follows: 
Panel (A): 0.3 (circles), 0.8 (squares), 1 (diamonds), 
1.25 (up triangles), 1.48 (left triangles),
and 2 (down triangles); and
Panel (B): 0.5 (circles), 1 (squares), 1.5 (diamonds), 
2 (up triangles), 3 (left triangles), and
4 (down triangles). Solid lines are guides. 
The insets show the low current resistivities. Currents,resistivities are 
measured in units of $J_{0}, \rho_{BS}$ respectively, as defined in the text.
Color on-line.}
\label{fig1}
\end{figure}

Our results for the resistivity $\rho$ versus the applied current
density $J$ are shown for a series of vortex 
densities below and above the matching field $B_{\Phi}$
for two different values of $B_{\Phi}$ in Fig. (\ref{fig1}A and B).
All error bars are equal to the symbol size. For the pinned Bose-glass 
regime $B < B_{\Phi}$ and for appreciable $B_{\Phi}$ 
(Fig. 1A), vortices are localized on separate columnar defects for 
small $J$ until an abrupt transition to a moving regime ensues 
near $J\sim 0.1 J_{0}$where the vortices simultaneously become unpinned. 
For larger driving currents, vortices are in the flux flow regime and the
resistivity approaches $\rho_{BS}$. 
The de-pinning transition occurs within a narrow range of currents 
corresponding to single vortex pinning. Consequently, as $B$ 
increases,  
the depinning transition occurs for smaller $J$ and significantly broadens
as the effective role of disorder diminishes relative
to the vortex lattice interaction, 
in agreement with previous simulations\cite{Trivedi}. 
The resistivity increases monotonically with increasing $B$ as 
shown in the inset of Fig. (1A). 
For values of $B$ near and above the matching field,
a low-current tail develops for small $J$ and continues to increase as
more and more interstitial vortices are not
able to reside on columnar defects due to the repulsion from the
pinned vortices. These interstitial vortices are caged by the
pinned vortices forming an weakly pinned interstitial glass 
with a much reduced critical current $J_{c}$, as suggested in
Refs. \cite{rad,LV}. As $B$ is further increased, the interstitial glass
melts into a interstitial liquid where channels
of vortices flow around pinned vortices for arbitrarily small $J$
and give a resistivity proportional to
the fraction of interstitial to pinned vortices. 
We note that for our parameter choices we do not
observe a Mott insulating phase as we always have interstitial
vortices for $B/B_{\Phi}=1$ due to the long-range vortex repulsion and high
temperatures.

\begin{figure}
\psfig{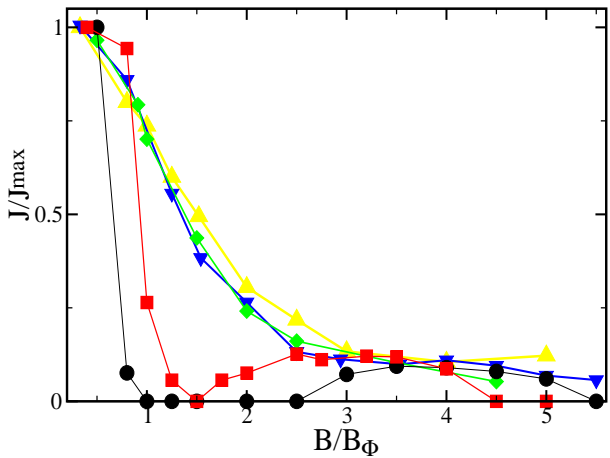} \caption[] 
{Normalized critical currents as a function of $B/B_{\Phi}$ for different
densities of columnar defects. The density of defects is given by $10^{3}\times B_{\Phi}/H_{c2}=1.525,
1.906, 3.05, 4.88,$ and $7.32$
for circles, squares, down triangles, 
diamonds, and up triangles, respectively. Color on-line.} \label{fig2}
\end{figure}

Important differences however are observed for smaller values of 
$B_{\Phi}$ (Fig. 1B). For $B<B_{\Phi}$ in this case, the transition to 
depinning occurs at lower $J$ and is substantially
broadened compared to Fig. (1A). A non-monotonic dependence of the 
resistivity appears as shown in the inset of Fig. (1B). For low driving 
forces the resistivity rises with increasing $B$ for $B/B_{\Phi}<1.5$ 
and becomes greater than $0.1\rho_{BS}$ for $B>B_{\Phi}$ as the 
interstitial glass melts into an interstitial liquid as the
density of vortices is increased. However at larger fields the 
resistivity decreases and
becomes less than $0.1\rho_{BS}$ for fields near three times the 
matching field.{\it This indicates that for low $B_{\Phi}$
the Bose glass melts into an interstitial
liquid near the matching field and recrystalizes into an
interstitial glass at larger values of $B/B_{\Phi}$, 
in contradiction to the phase diagram proposed previously\cite{rad}}. 

\begin{figure}
\psfig{file=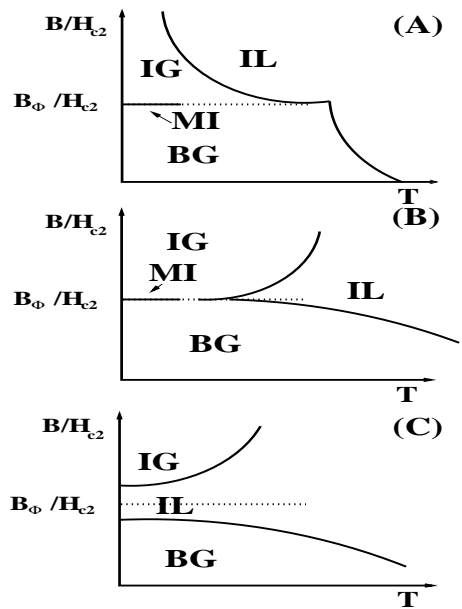,width=6cm,height=8cm} \caption[]
{Conjectured phase diagram for large (A), intermediate (B), and small
(C) matching field densities.} \label{fig3}
\end{figure}

We can make this more quantitative by defining a critical 
current density $J_{c}$ as the value of the current density 
corresponding to a resistivity of 10\% of $\rho_{BS}$. 
For a number of different molecular dynamics runs of different $B_{\Phi}$,
the values we obtain for $J_{c}$ are shown in Fig. (2) as a function of 
$B/B_{\Phi}$ normalized to the values of $J_{c}$ determined in the 
small $B/B_{\Phi}$ limit. For large $B_{\Phi}$, the critical current 
decreases monotonically with $B/B_{\Phi}$ with a gradual fall-off
near the matching field. For smaller values of $B_{\Phi}$ the fall-off 
near $B_{\Phi}$ becomes much more abrupt, suggesting that a magnitude of
the kink in the melting curve would be dependent on the value of $B_{\Phi}$,
reconciling previous 
experiments\cite{Beauchamp,Nowak,worth,Krusin,Nakielski,Samoilov,Maz,Smith99,Smith01}.
For even smaller $B_{\Phi}$, the fall-off is dramatic and
the interstitial glass weakens appreciably and melts ($J_{c}=0$) 
near a critical value of $B_{\Phi}^{c}/H_{c2}\sim 1.9\times 10^{-3}$. Remarkably,
the critical current resurrects for larger values of the magnetic field and has a broad peak
near $B=3B_{\Phi}$ for $B_{\Phi}/H_{c2}\times 10^{3}=1.906$
before falling off again at still larger fields. This reproduces the peak near
three times the matching field seen in experiments on Tl$_{2}$Ba$_{2}$CuO$_{6+\delta}$
\cite{Nowak}. However we see that still smaller values of $B_{\Phi}$ yields a critical
current peak at larger values of $B/B_{\Phi}$ suggesting that this peak would be
dependent on the number of defects. While we expect that the
actual value of the critical field $B^{c}$ might depend on our choice of defining $J_{c}$,
we do not expect the rentrant behavior into the interstitial glass to be qualitativelty
changed. A proposed phase diagram which encompasses our results for different $B_{\Phi}$
is given in Figure (\ref{fig3}).

Using renormalization
group arguments it was shown by Nelson and Seung that for clean systems
the interactions between vortex
lines are renormalized to zero near $H_{c1}$ and leads to the melting
of the Abrikosov lattice as $H\rightarrow H_{c1}$\cite{NS}. Thus we might expect that
the melting of the interstitial glass for $B-B_{\Phi}\ge 0^{+}$ is similar to the
melting of the vortex lattice near $H_{c1}$\cite{Nelson1,NS}. However for $B<B_{\Phi}$
vortex localization onto unoccupied columnar defects helps to reduce vortex fluctuations
and one might expect the Bose glass to be stable near $H_{c1}$ as vortices occupy the
strongest defects. Yet once $B>B_{\Phi}$ the interstitial vortices feel a much reduced
and screened defect interaction and suffer the downward renormalization of the shear
modulus as in clean systems. This is borne out in our simulations for small defect
concentrations. As the field increases 
our simulations indicate that the interstitial glass recrystalizes. 
This can be viewed as a hardening of the shear modulus for increasing fields 
as the role of interactions increases.

What leads to the recrystalization of the melted interstitials at larger values of $B$ for small defect
concentrations? At large vortex concentrations the Coulomb potential is effectively short-range due
to screening and the effects of disorder are also screened by flux line collisions, and indeed our
results are consistent with prior simulations for contact
repulsive potentials\cite{Trivedi}. The system favors forming dislocations at larger fields and the
domains are vortices are easily depinned. For smaller vortex
concentrations, new physics arises as screening is no longer effective and the bare long-range repulsive
forces encourage longer-range ordering concommitant with decreased prevalence of dislocation and
subsequent enhanced vortex pinning.

\begin{figure}
\psfig{file=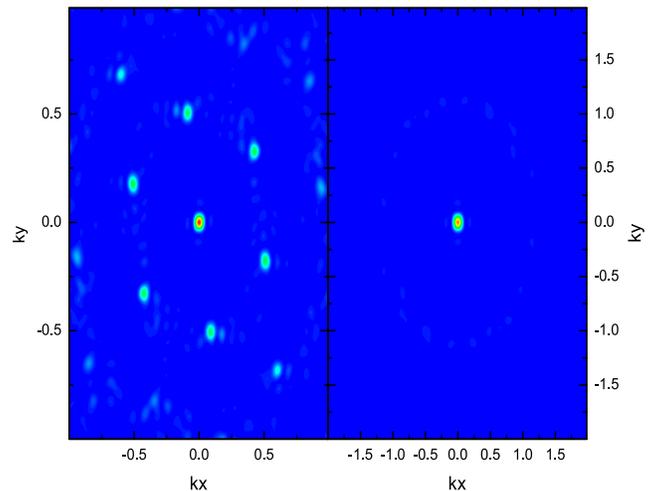,width=9cm,height=7cm} \caption[] {
Structure factor in the pinned phase at three times the matching field 
$B/B_{\Phi}=3$ for $10^{3}\times B_{\Phi}/H_{c2}=1.525$ (left) and
7.32 (right) panels, respectively. Momentum is measured in units of $1/d$.
Color on-line.} \label{fig4}
\end{figure}

To check these ideas, we plot in Figure (\ref{fig4}) time-averaged
structure plots for $B=3B_{\Phi}$ at two values of $B_{\Phi}$.
While long-range order is not seen even on the length scale of our simulations, it is clear
that orientational short-range order is more prevalent for the small $B_{\Phi}$ results than the large
ones, with subsidiary structure peaks at the reciprocal lattice vectors up to 1/2 the height of
the central ($Q=0$) peak. When averaged over many disorder configurations, the orientational order
is lost for both values of $B_{\Phi}$, yet it is clear that the effective range of the vortex
repulsion is longer for smaller $B_{\Phi}$. We would expect that for still smaller 
values of $B_{\Phi}$ positional order would grow for systems of fixed finite size $L$
with the possible formation of Bragg-like peaks. Most likely the positional order would not be
quantitatively relevant for $L\rightarrow \infty$.

In summary, we have presented numerical simulations of interacting
vortex dynamics in the presence of columnar disorder as a function of $B/B_{\Phi}$
and $B_{\Phi}/H_{c2}$. We find a monotonic decrease in the critical current for large values of $B_{\Phi}$,a
with a sharp drop-off near the matching field, consistent
with prior notions of the weakly pinned intersitial glass emerging at higher fields. For smaller values of
$B_{\Phi}$ however, we see abrupt melting at $B=B_{\Phi}$ and recrystalization near $B=3B_{\Phi}$
of the interstitial glass, suggesting modifications to the phase diagram due to the role of
long-range interactions even for columnar disorder.

\acknowledgements {T.P.D. would like to thank T. Giamarchi, M. J.
P. Gingras, and T. F. Rosenbaum for many useful discussions.
T.P.D. would like to acknowledge support from NSERC, PREA,
Canadian Foundation for Innovation, the Ontario Innovation Trust,
and the Alexander von Humboldt foundation.}

\end{document}